\documentclass[
reprint,
amsmath,amssymb,
aps,
prc,
floatfix,
]{revtex4-2}

\usepackage{graphicx}
\usepackage{dcolumn}
\usepackage{bm}
\usepackage{diagbox}
\usepackage{tabularx}
\usepackage{booktabs} 
\usepackage{amsmath}
\usepackage{epstopdf}
\usepackage{hyperref}[colorlinks, linkcolor=blue, anchorcolor=blue, citecolor=blue]
\usepackage[mathlines]{lineno}

\begin{document}


\title{Production cross sections of superheavy elements: insights from the dinuclear system model with high-quality microscopic nuclear masses}

\author{Peng-Hui Chen}%
 \email{chenpenghui@yzu.edu.cn}
\affiliation{School of Physical Science and Technology, Yangzhou University, Yangzhou 225009, China}%

\author{Chang Geng}
 \affiliation{School of Physical Science and Technology, Yangzhou University, Yangzhou 225009, China} 

\author{Zu-Xing Yang}
\affiliation{RIKEN Nishina Center, Wako, Saitama 351-0198, Japan}

\author{Xiang-Hua Zeng}
\affiliation{School of Physical Science and Technology, Yangzhou University, Yangzhou 225009, China}

\author{Zhao-Qing Feng}
\email{fengzhq@scut.edu.cn}
\affiliation{School of Physics and Optoelectronics, South China University of Technology, Guangzhou 510641, China}%

\date{\today}

\begin{abstract}
To accurately predict the synthesis cross-sections of superheavy elements, identifying the optimal projectile-target combinations and the evaporation channels at specific collision energies, we have attempted to utilize high-quality microscopic nuclear masses (HQMNM) within the dinuclear system (DNS) model, which are obtained by fitting experimental data with the Skyrme energy density functional theory (DFT), as published in Phys. Lett. B 851 (2024) 138578. The atomic nuclear mass serves as a crucial input for the DNS model, as the Q-values and separation energies it generates directly influence the calculated fusion and survival probabilities. Our calculations have reproduced the experimental data for hot fusion and have been compared with results based on the finite-range droplet model (FRDM12) mass calculations. Compared to the FRDM12 mass results, we have found that the HQMNM provides a better fit to the experimental outcomes. 
For the specific reaction of \(^{48}\rm{Ca} + ^{243}\rm{Am} \rightarrow ^{291}\rm{Mc}^*\), we have conducted a detailed calculation of capture, fusion, and survival based on the HQMNM model and compared these with calculations based on other mass models. Based on these findings, we have systematically calculated available projectile target combinations for the synthesis of elements 119 and 120, and identified the optimal combinations. We provided the synthesis cross-sections, collision energies, and evaporation channels, offering a reference for conducting experiments on the synthesis of superheavy elements.
\end{abstract}

\maketitle


\section{Introduction}\label{sec1}

Currently, the synthesis of superheavy elements (SHE) with atomic numbers 119 and 120 represents the frontier of research in nuclear physics\cite{PhysRevC.79.024603,PhysRevC.102.064602,Hofmann2016}. The creation of these elements is pivotal to the initiation of the eighth period of the periodic table and the expansion of the nuclear chart. The synthesis of superheavy nuclei contributes to the study of the limits of nuclear existence and the properties of nuclear and atomic structures under such extreme Coulomb force conditions, serving to validate existing theoretical models. The fission of superheavy nuclei is crucial for astrophysical research on the rapid neutron-capture process (r-process)\cite{RevModPhys.93.015002,Hao2022}. Presently, the experimental synthesis of superheavy nuclei is primarily based on fusion-evaporation reactions. Over the past few decades, significant progress has been made in the synthesis of superheavy elements, with experiments employing cold or hot fusion reactions to create elements from 104 to 118\cite{Hof96112,doi:10.1143/JPSJ.73.2593,PhysRevLett.105.182701,PhysRevC.69.021601,PhysRevC.69.054607,PhysRevLett.104.142502,PhysRevC.74.044602}. 
To synthesize unknown SHEs, China, Russia, Japan, and Germany have constructed heavy-ion accelerators and spectrometer facilities, such as the CAFE2 and SHANS2 in Lanzhou, China\cite{Gan2022}; the DGFRS in Dubna, Russia\cite{OGANESSIAN2022166640}; the GAERIS at RIKEN, Japan\cite{MORITA201530}; and the superSHIP at GSI, Germany\cite{Mnzenberg2023}. Each country has also formulated plans for the synthesis of superheavy elements.
The synthesis of SHE is challenging because of limitations in projectile-target materials and extremely low synthesis cross sections, resulting in long experimental timelines and high costs. Therefore, it is essential to systematically calculate the excitation functions of the evaporation residue cross sections for superheavy nuclei synthesis using theoretical models before conducting experiments. This provides experimentalists with optimal projectile-target combinations, the most suitable beam energies, and synthesis cross sections for reference. There are various theoretical models available for studying the synthesis mechanism of superheavy nuclei, such as the time-dependent Hartree-Fock (TDHF) model \cite{GUO2018401,10.3389/fphy.2019.00020,MARUHN20142195}, the improved quantum molecular dynamics models (ImQMD) \cite{PhysRevC.65.064608,PhysRevC.88.044611,PhysRevC.88.044605}, the multidimensional Langevin equations \cite{PhysRevC.96.024618,PhysRevC.85.014608}, and the dinuclear system (DNS) model \cite{FENG200650,PhysRevC.91.011603,PhysRevC.89.024615,epja20gga,07fengcpl}. This article will employ the DNS model to predict the synthesis cross-sections of superheavy nuclei. The DNS model offers a clear physical picture, more closely resembling the actual process of heavy-ion collisions, and can self-consistently account for important factors such as dynamic energy dissipation, nucleon transfer, fission, quasi-fission, shell effects, pairing effects, and dynamic deformation\cite{PhysRevC.91.011603,PhysRevC.93.064610,2023chen,niu2020,2018chen,Chen20232,PhysRevC.68.034601,PhysRevC.106.054601}. However, the calculations based on the DNS model are highly dependent on basic nuclear data, such as nuclear masses, deformations, shell and pairing corrections, interaction potentials, Coulomb barriers, and fission barriers. This paper will primarily investigate the dependence of the DNS model's calculation results on nuclear masses.

Based on the DNS model, the values $Q$, the binding energies and the neutron separation energies derived from the nuclear masses directly influence the calculations of the capture cross sections, the probability of fusion and the probability of survival. Nuclear masses encompass a wealth of information about the nucleus, such as symmetry energy, pairing corrections, shell corrections, and deformation energy, which can affect the accuracy of predicting superheavy elements. Moreover, there are certain differences in the mass data tables provided by different nuclear mass models, especially for unknown nuclei.
Our previous work has discussed the impact of nuclear masses on calculations of fusion probability \cite{PhysRevC.109.054611}. 
This paper will focus on the use of a high-quality microscopic nuclear mass (HQMNM)\cite{GUAN2024138578} to predict the synthesis cross sections of new superheavy elements within the DNS model and compare the results with those of four other nuclear mass models, such as FRDM12\cite{MOLLER20161}, WS4\cite{WANG2014215}, KUTY\cite{10.1143/PTP.113.305} and HFB02\cite{ELBASSEM201722}.

\section{Model Description}\label{sec2}

Based on the DNS model, the evaporation residual cross-section of SHE can be expressed as
\begin{eqnarray}
\sigma_{\mathrm{ER}}\left(E_{\mathrm{c} . \mathrm{m} .}\right)  =  \frac{\pi \hbar^{2}}{2 \mu E_{\mathrm{c} . \mathrm{m} .}} \sum_{J  =  0}^{J_{\max }}(2 J+1) T\left(E_{\mathrm{c} . \mathrm{m} .}, J\right) \nonumber \\
\times P_{\mathrm{CN}}\left(E_{\mathrm{c} . \mathrm{m} .}, J\right) W_{\mathrm{sur}}\left(E_{\mathrm{c} . \mathrm{m} .}, J\right).
\end{eqnarray}
Here, the $ T\left(E_{\mathrm{c} . \mathrm{m} .}, J\right) $ is penetration probability, which is the probability of the colliding system overcoming the Coulomb barrier \cite{FENG200650}. The $ P_{\mathrm{CN}}\left(E_{\mathrm{c} . \mathrm{m} .}, J\right) $ is the fusion probability \cite{PhysRevC.76.044606,PhysRevC.80.057601}.
The $ W_{\mathrm{sur}}\left(E_{\mathrm{c} . \mathrm{m} .}, J\right) $is the survival probability of the highly excited compound nuclei, which survived by evaporating light particles against fission. 

The $T\left(E_{\mathrm{c} . \mathrm{m} .}, J\right)$ is derived by Hill-Wheeler formula\cite{PhysRev.89.1102} with barrier distribution function \cite{PhysRevC.101.024610}, written as
\begin{eqnarray}\label{hwl}
&T(E_{\mathrm{c.m.}},J)=\displaystyle\int \frac{f(B) \mathrm{d}B}{1+\exp\Big\{-\frac{2\pi (E_{\mathrm{c.m.}}-B-E_{\rm rot})}{\hbar\omega(J)}\Big\}}.
\end{eqnarray}
Here $\hbar \omega (J)$ is the width of the parabolic barrier at $R_{B}(J)$. The normalization constant is $\int f(B)dB=1$.
The Coulomb barrier distribution is taken as the asymmetry Gaussian formula \cite{PhysRevC.65.014607,Chen_2017}.
The nucleus-nucleus potential is calculated by the double-folding method  \cite{PhysRevC.80.057601,PhysRevC.76.044606, J.Mod.Phys.E5191(1996)}.
The Coulomb potential is evaluated by Wong's formula \cite{PhysRevLett.31.766}.


In the fusion stage, the formation probabilities of fragments P($Z_{1},N_{1},E_{1}$) are estimated by solving a set of master equations \cite{FENG201082,FENG200933,PhysRevC.76.044606}, 
\begin{eqnarray}
&&\frac{d P(Z_1,N_1,E_1,t)}{d t} = \nonumber \\ && \sum \limits_{Z'_1}W_{Z_1,N_1;Z_1,N'_1}(t) [d_{Z_1,N_1}P(Z'_1,N_1,E'_1,t) \nonumber \\ && - d_{Z'_1,N_1}P(Z_1,N_1,E_1,t)] + \nonumber \\ &&
 \sum \limits_{N'_1}W_{Z_1,N_1;Z_1,N'_1}(t)[d_{Z_1,N_1}P(Z'_1,N_1,E'_1,t) \nonumber \\ && - d_{Z_1,N'_1}P(Z_1,N_1,E_1,t)] - \nonumber \\
 &&[\Lambda ^{\rm qf}_{A_1,E_1,t}(\Theta) + \Lambda^{\rm fis}_{A_1,E_1,t}(\Theta)]P(Z_1,N_1,E_1,t).
\end{eqnarray}
Here, the $W_{\rm Z_1,N_1,Z'_1,N_1}$ is the mean transition probability from the channel ($Z_1,N_1,E_1$) to ($Z'_1,N_1,E'_1$), and $d_{\rm Z_1,N_1}$ denotes the microscopic dimension corresponding to the macroscopic state ($Z_1,N_1,E_1$).
The $E_1$ is the local excitation energy $\varepsilon^*_1$ of fragment $A_1$ \cite{PhysRevC.27.590}. The sticking time is derived by the parametrization method of classical deflection function \cite{LI1981107}. 
The quasi-fission evaluated by Kramers equation \cite{PhysRevC.68.034601,PhysRevC.27.2063}.

The potential energy surface (PES) of the DNS is given by
\begin{eqnarray}
&&U_{\rm dr}(\{\alpha\})=B_1+B_2-B_{\rm CN}+V_{\rm CN}(\alpha)+E_{\rm rot} 
\end{eqnarray}
Here, the $\{\alpha\}$ stands for $\{r,\beta_{1},\beta_{2},\theta _{1},\theta_{2}\}$. $B_{\rm i}$ ($\rm i$ = 1, 2) and $B_{\rm CN}$ are the negative binding energies of the fragment $A_{\rm i}$ and the compound nucleus $A_{\rm CN}$, respectively, which are taken from nuclear mass tables. The $E_{\rm rot}$ is the rotation energy of the compound nucleus. 

The DNS model posits that all fragments capable of surpassing the Businaro-Gallone (BG) point can undergo fusion. Hence, the fusion probability can be written as 
\begin{eqnarray}
P_{\rm CN}(E_{\rm c.m.},V_{\rm in})=\sum _{Z=1}^{Z _{\rm BG}}\sum_{N=1}^{N_{\rm BG}} P(Z,N,E',V_{\rm in}).
\end{eqnarray}
Here $Z _{\rm BG}$ and $N_{\rm BG}$ are the proton number and neutron number of the BG point. The $V_{\rm in}$ is the potential energy at the initial contact point.

The excited compound nuclei would be de-excited by evaporating $\gamma$-rays, neutrons, protons, $\alpha$ $etc.$) against fission. The survival probability of the evaporation channels x-th neutron, y-th proton, and z-th alpha are evaluated by a statistical evaporation model based on Weisskopf's evaporation theory, expressed as \cite{Chen2016,FENG201082,FENG200933,Xin2021}
\begin{eqnarray}
&&W_{\rm sur}(E_{\rm CN}^*,x,y,z,J)=P(E_{\rm CN}^*,x,y,z,J) \times \nonumber\\&&  \prod_{i=1}^{x}\frac{\Gamma _n(E_i^*,J)}{\Gamma _{\rm tot}(E_i^*,J)} \prod_{j=1}^{y}\frac{\Gamma _p(E_j^*,J)}{\Gamma_{\rm tot}(E_i^*,J)} \prod_{k=1}^{z}\frac{\Gamma _\alpha (E_k^*,J)}{\Gamma _{\rm tot}(E_k^*,J)}, 
\end{eqnarray}
where the $E_{\rm CN}^*$ and $J$ are the excitation energies and the spin of the compound nuclei, respectively. The total width $\Gamma_{\rm tot}$ is the sum of partial widths of emitting particles, $\gamma$-rays, and fission. 
The $P(E_{\rm CN}^*, J)$ is the realization probability of evaporation channels.  
The $\Gamma _n(E_i^*,J)$, $\Gamma _p(E_i^*,J)$ and $\Gamma _{\alpha}(E_i^*,J)$ are the decay widths of particles n, p, $\alpha$, which are evaluated by the single-particle level density \cite{PhysRevC.68.014616}. The fission width $\Gamma_f(E^*,J)$ is calculated by the Bohr-Wheeler formula \cite{PhysRevC.80.057601,artza05,PhysRevC.76.044606}.
The fission barrier utilized in this study is derived from the macroscopic-microscopic model. \cite{MYERS19661,MOLLER20161}. 

\section{\label{sec3} Results and discussions}

\begin{table}[h]
\renewcommand{\arraystretch}{1.1}
\setlength{\tabcolsep}{2.5pt}
\caption{$Q$-value of the reactions based on three nuclear mass models and their Bass potential energies.}
\label{table1}
\begin{tabular}{lllll}
\hline
Reactions $\backslash$ Mass & HQMNM & FRDM & WS4  & $V_{\rm B}$\\
\hline
$^{48}$Ca+$^{238}$U$\rightarrow$ $^{286}$Cn$^*$ & -159.42 & -157.76 & -160.4  & 191.43\\
$^{48}$Ca+$^{237}$Np$\rightarrow$ $^{237}$Np$^*$ & -162.53 & -161.83 & -164.54 & 193.85\\
$^{48}$Ca+ $^{242}$Pu$\rightarrow$ $^{290}$FI$^*$ & -163.38 & -161.84 & -165.12 &195.15\\
$^{48}$Ca+$^{244}$Pu$\rightarrow$ $^{292}$FI$^*$ & -160.44 & -160.44 & -163.7   & 194.78\\
$^{48}$Ca+$^{243}$Am$\rightarrow$ $^{291}$Mc$^*$ & -165.97 & -164.98 & -168.24   & 197.19\\
$^{48}$Ca+$^{245}$Cm$\rightarrow$ $^{293}$Lv$^*$ & -168.26 & -167.81 & -170.93 & 199.03\\
$^{48}$Ca+ $^{248}$Cm$\rightarrow$ $^{296}$Lv$^*$ & -166.57 & -166.57 & -169.49  &198.48\\
$^{48}$Ca+$^{249}$Bk$\rightarrow$ $^{297}$Ts$^*$ & -170.07 & -170.07 & -172.87  & 200.5\\
$^{48}$Ca+$^{249}$Cf$\rightarrow$ $^{297}$Og$^*$& -174.18 & -174.18 & -176.67  & 202.73\\
$^{58}$Fe+$^{237}$Np$\rightarrow$$^{295}$119$^*$ & -220.25  & -219.6 & -222.67  & 249.7\\
$^{54}$Cr+$^{243}$Am$\rightarrow$$^{297}$119$^*$ & -204.88 & -204.65 & -207.32 & 235.63\\
$^{50}$Ti+$^{249}$Bk$\rightarrow$$^{299}$119$^*$ & -190.14 & -190.66 & -191.97 & 220.77\\
$^{51}$V+$^{248}$Cm$\rightarrow$$^{299}$119$^*$ & -193.34 & -193.86 & -195.4 & 228.56\\
$^{55}$Mn+$^{244}$Pu$\rightarrow$$^{299}$119$^*$ & -205.15 & -205.67 & -208.5 & 242.5\\
$^{55}$Mn+$^{243}$Am$\rightarrow$$^{298}$120$^*$ & -211.85 & -211.36 & -213.74  & 245.54\\
$^{50}$Ti+$^{249}$Cf$\rightarrow$$^{299}$120$^*$ & -196.39 & -195.97 & -196.93 & 223.22\\
$^{51}$V+$^{249}$Bk$\rightarrow$$^{300}$120$^*$ & -178.02 & -178.02 & -180.85 & 209.38\\
$^{54}$Cr+$^{248}$Cm$\rightarrow$$^{302}$120$^*$ & -205.62 & -206385 & -208.63 & 237.18\\
$^{58}$Fe+$^{244}$Pu$\rightarrow$$^{302}$120$^*$ & -218.04 & -219.27 & -221.31 & 250.89\\
\hline
\end{tabular}
\end{table}

\begin{table*}[htb]
\renewcommand{\arraystretch}{1.2}
\setlength{\tabcolsep}{6pt}
\caption{Predictions of maximum production cross sections of SHE via HQMNM, compared to other nuclear mass models of FRDM12, WS4, KUTY, and HFB, based on the dinuclear system model.}
\label{table2}
\begin{tabular}{lrrrrrr}
\hline
Reactions $\backslash$ Mass & HQMNM\cite{GUAN2024138578} & FRDM12\cite{MOLLER20161} & WS4\cite{WANG2014215}  & KUTY\cite{10.1143/PTP.113.305} & HFB02\cite{ELBASSEM201722} & EXP \\
 A$_{\rm P}$(A$_{\rm T}$, xn)A$_{\rm CN}$ & $\sigma^{\rm CAL}_{\rm max}$, $E^*$(MeV) & $\sigma^{\rm CAL}_{\rm max}$ $(E^*)$ & $\sigma^{\rm CAL}_{\rm max}$ $(E^*)$  & $\sigma^{\rm CAL}_{\rm max}$ $(E^*)$ & $\sigma^{\rm CAL}_{\rm max}$ $(E^*)$ & $\sigma^{\rm EXP}_{\rm max}$ $(E^*)$ \\
\hline
$^{48}$Ca($^{238}$U,3n)$^{283}$Cn    & 2.85 pb (34) & 1 pb (36) & 0.5 pb (34)  & 0.4  pb (32) & 0.87 pb (32) & 0.74 pb (34)  \cite{Hof96112}\\
$^{48}$Ca($^{237}$Np,3n)$^{283}$Nh   & 1.36 pb (34) & 1.43 pb (36) & 0.3  pb (34)  & 0.2  pb (34) & 0.09 pb (36) & 0.87 pb (39) \cite{doi:10.1143/JPSJ.73.2593}\\
$^{48}$Ca($^{242}$Pu,4n)$^{287}$Fl   & 2.67 pb (38) & 0.68 pb (40) & 0.6  pb (40)  & 0.2  pb (40) & 0.57 pb (40) & 4.49 pb (40) \cite{PhysRevLett.105.182701}\\
$^{48}$Ca($^{244}$Pu,4n)$^{288}$Fl   & 1.97 pb (38) & 3.3  pb (38) & 1    pb (38)  & 0.2  pb (40) & 1.11 pb (38) & 10 pb (42) \cite{PhysRevLett.105.182701}\\
$^{48}$Ca($^{243}$Am,3n)$^{283}$Mc   & 17   pb (30) & 5.39 pb (32) & 18.8 pb (34)  & 0.7  pb (32) & 12.6 pb (34) & 17.5 pb (34) \cite{PhysRevC.69.021601}\\
$^{48}$Ca($^{245}$Cm,3n)$^{283}$Lv   & 3.37 pb (32) & 1.99 pb (32) & 1.1  pb (32)  & 0.37 pb (32) & 0.9  pb (32) & 3.67 pb (38) \cite{PhysRevC.69.054607}\\
$^{48}$Ca($^{248}$Cm,4n)$^{287}$Lv   & 1.51 pb (38) & 1.43 pb (38) & 47.5 pb (42)  & 0.08 pb (40) & 0.35 pb (38) & 3.41 pb (40) \cite{PhysRevC.69.054607}\\
$^{48}$Ca($^{249}$Bk,4n)$^{288}$Ts   & 0.8  pb (32) & 2.36 pb (30) & 0.12 pb (38)  & 4.8  fb (36) & 20.6 fb (36) & 2.49 pb (42) \cite{PhysRevLett.104.142502}\\
$^{48}$Ca($^{249}$Cf,3n)$^{288}$Og   & 0.2  pb (34) & 0.33 pb (34) & 31.5 pb (34)  & 9.79 fb (36) & 18.3 fb (34) & 0.56 pb (34) \cite{PhysRevC.74.044602}\\
$^{58}$Fe($^{237}$Np,3n)$^{292}$119  & 23.7 fb (36) & 2.8  fb (36) & 1.13 fb (34)  & 0.19 fb (34) & 0.2fb (34) & --\\
$^{54}$Cr($^{243}$Am,3n)$^{294}$119  & 4.64 fb (34) & 7.6  fb (34) & 1.28 fb (34)  & 0.2  fb (36) & 0.47 fb (34) & --\\
$^{50}$Ti($^{249}$Bk,3n)$^{296}$119  & 5.16 fb (36) & 12.6 fb (34) & 6.73 fb (32)  & 1.08 fb (34) & 0.58 fb (34) & -- \\
$^{51}$V($^{248}$Cm,3n)$^{296}$119   & 3.63 fb (36) & 8.1  fb (36) & 6.1  fb (34)  & 1.29 fb (34) & 7.4 fb (34) & -- \\
$^{55}$Mn($^{244}$Pu,3n)$^{296}$119  & 3.07 fb (36) & 7.73 fb (32) & 4.55 fb (32)  & 0.6 fb (34)  & 0.27 fb (34) & -- \\
$^{55}$Mn($^{243}$Am,3n)$^{295}$120  & 2.19 fb (34) & 0.9  fb (36) & 0.05 fb (38)  & 0.007 fb (40)& 0.04 fb (36) & -- \\
$^{50}$Ti($^{249}$Cf,3n)$^{296}$120  & 4.35 fb (34) & 8.8 fb (34)  & 3.27 fb (36)  & 0.63 fb (40) & 1.25 fb (36) & -- \\
$^{51}$V($^{249}$Bk,3n)$^{297}$120   & 1.78 fb (36) & 1.3 fb (36)  & 3.05 fb (34)  & 0.65 fb (36) & 6.98 fb (34) & -- \\
$^{54}$Cr($^{248}$Cm,3n)$^{299}$120 & 0.09 fb (34)  & 0.19 fb (34) & 0.09 fb (34)  & 0.009 fb (38)& 0.08 fb (38) & -- \\
$^{58}$Fe($^{244}$Pu,3n)$^{299}$120  & 0.17 fb (34) & 0.36 fb (34) & 0.24 fb (32)  & 0.02 fb (36) & 0.06 fb (38) & -- \\
\hline
\end{tabular}
\end{table*}

\begin{figure}[htb]
\includegraphics[width=1.\linewidth]{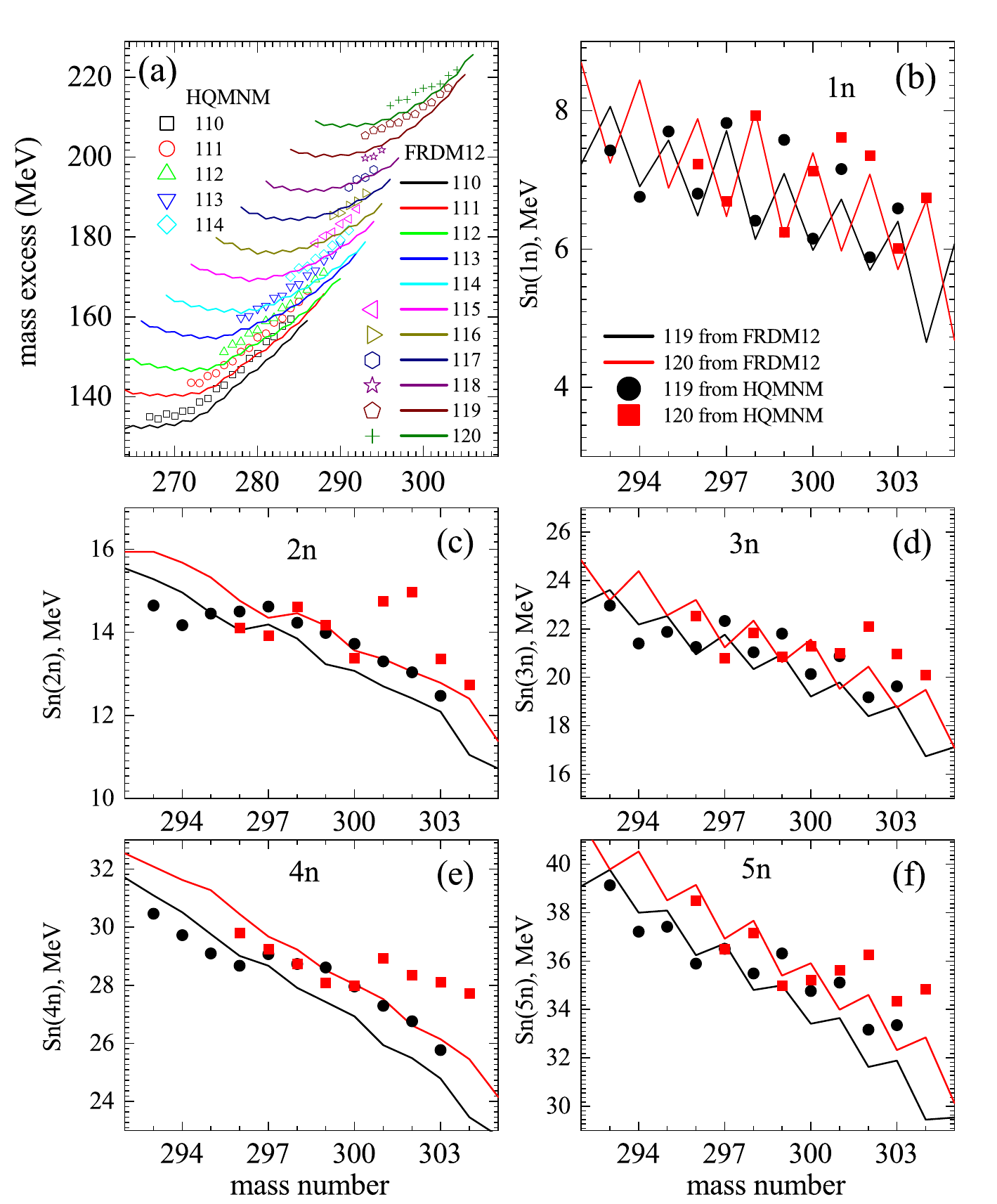}
\caption{\label{fig1} Panel (a) presents a comparison of the nuclear mass excess between the HQMNM and the FRDM12 for superheavy nuclei (SHN) with atomic numbers \( Z = 110-120 \). Panels (b) through (f) list the neutron separation energies for the 1-neutron to 5-neutron channels, respectively.}
\end{figure}

Figure \ref{fig1} presents the excess mass of HQMNM and FRDM12 and their neutron separation energies. Panel (a) lists the excess mass of SHN with atomic numbers 110 to 120, represented by discrete points and lines. The hollow black squares, red circles, green upward triangles, blue downward triangles, cyan diamonds, magenta left triangles, dark yellow right triangles, dark blue hexagons, purple stars, burgundy pentagons, and olive crosses (solid lines of the same color) represent isotopes of elements numbered 110 to 120, respectively. Panels (b) to (f) list the separation energies for 1n to 5n, respectively. The red and black solid lines represent the isotopic masses of elements 120 and 119 in the FRDM12 mass table, while the solid red squares and black dots represent the isotopic masses of elements 120 and 119 in the HQMNM mass table, respectively. It can be observed from panel (a) that the excess mass provided by the HQMNM mass table is generally greater than that of the FRDM12. There is a significant difference in the separation energies between the two mass tables. The separation energies for odd neutrons exhibit a pronounced odd-even effect. By comparing the excess masses of superheavy isotopes from the two mass tables, it is found that in the neutron-rich region, the separation energy provided by HQMNM is larger, while in the proton-rich region, the separation energy provided by FRDM12 is larger. The differences in separation energies between the two mass tables will be reflected in the survival probabilities of superheavy nuclei, and the differences in the absolute values of excess mass will be manifested in the fusion probabilities.
Table \ref{table1} lists the $Q$ values and Bass barriers for the reactions synthesizing elements 110 to 120, based on three mass tables of HQMNM, FRDM12, and WS4. It can be observed from the table that the differences provided by the different mass tables are within 3 MeV.

\begin{figure}[htb]
\includegraphics[width=0.85\linewidth]{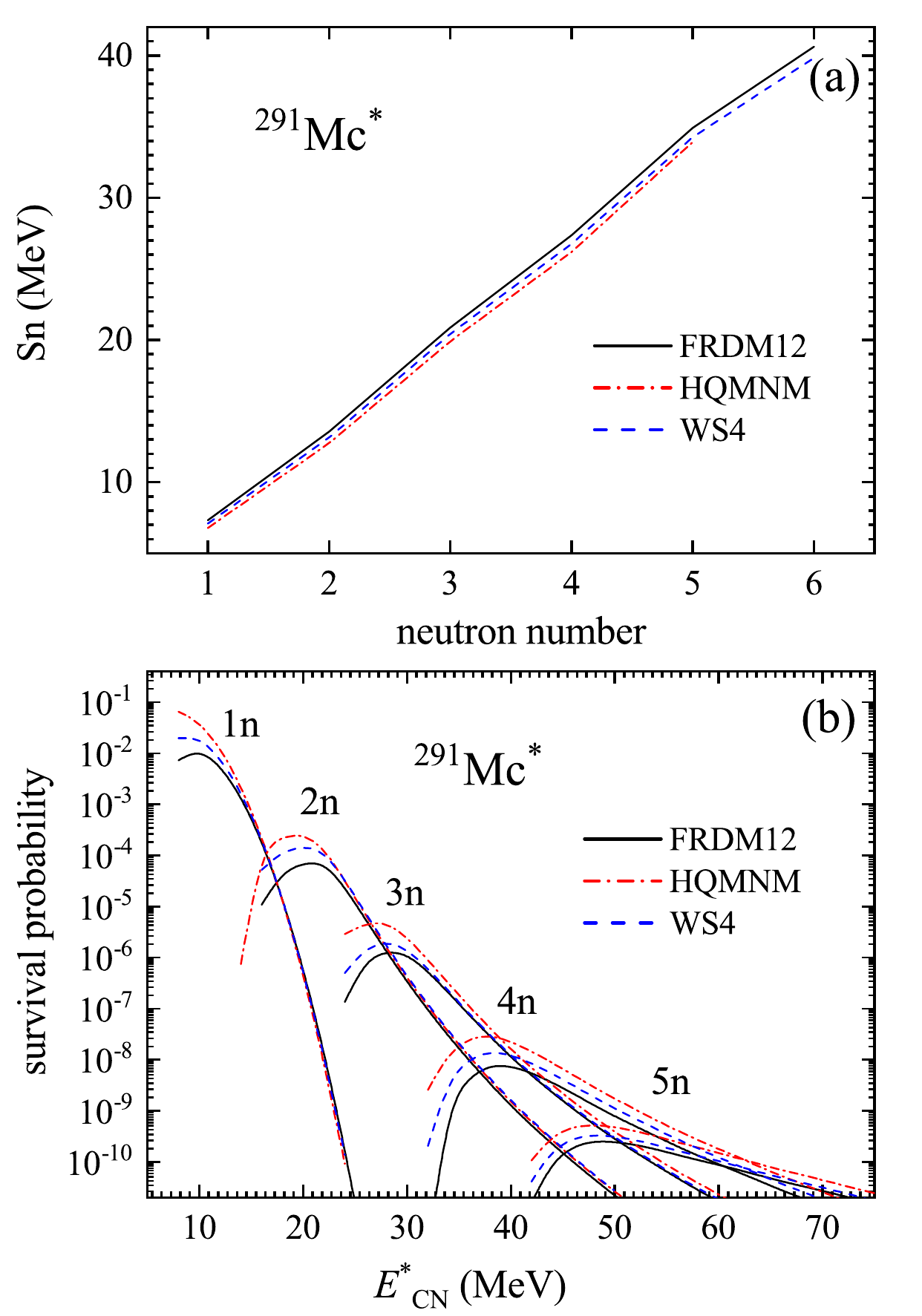}
\caption{\label{fig2} Panel (a) illustrates the separation energies of 1- to 5-neutrons for $^{291}$Mc. Panel (b) presents the survival probabilities of 1- to 5-neutron evaporation channels of the \(^{291}Mc^*\) have been calculated using the statistical evaporation model, based on the FRDM12, the HQMNM, and the WS4 model, respectively. The results are represented graphically by black solid lines for FRDM12, red dash-dot lines for HQMNM, and blue dashed lines for WS4, respectively.}
\end{figure}

Figure \ref{fig2} presents the neutron separation energies of $^{291}$Mc based on three mass tables of HQMNM, FRDM12, and WS4, along with the corresponding survival probabilities calculated using the statistical model. The black solid lines represent the results calculated based on the FRDM12 mass table, the blue dotted lines indicate the results based on the WS4 mass table, and the red dashed lines represent the results based on the HQMNM mass table. In panel (a), the distribution trend of the neutron separation energies for the superheavy nucleus $^{291}$Mc is ordered from largest to smallest as FRDM12, WS4, and HQMNM. The neutron separation energy represents the difficulty of neutron evaporation, which directly influences the calculation of survival probabilities. The larger the neutron separation energy, the smaller the survival probability for the neutron evaporation channel, and vice versa. Therefore, in panel (b), we can observe that the survival probabilities for the excited $^{291}$Mc$^*$ with neutron evaporation channels of 1n-5n are in the order based on the mass tables HQMNM, WS4, and FRDM12, respectively.

\begin{figure}[htb]
\includegraphics[width=0.99\linewidth]{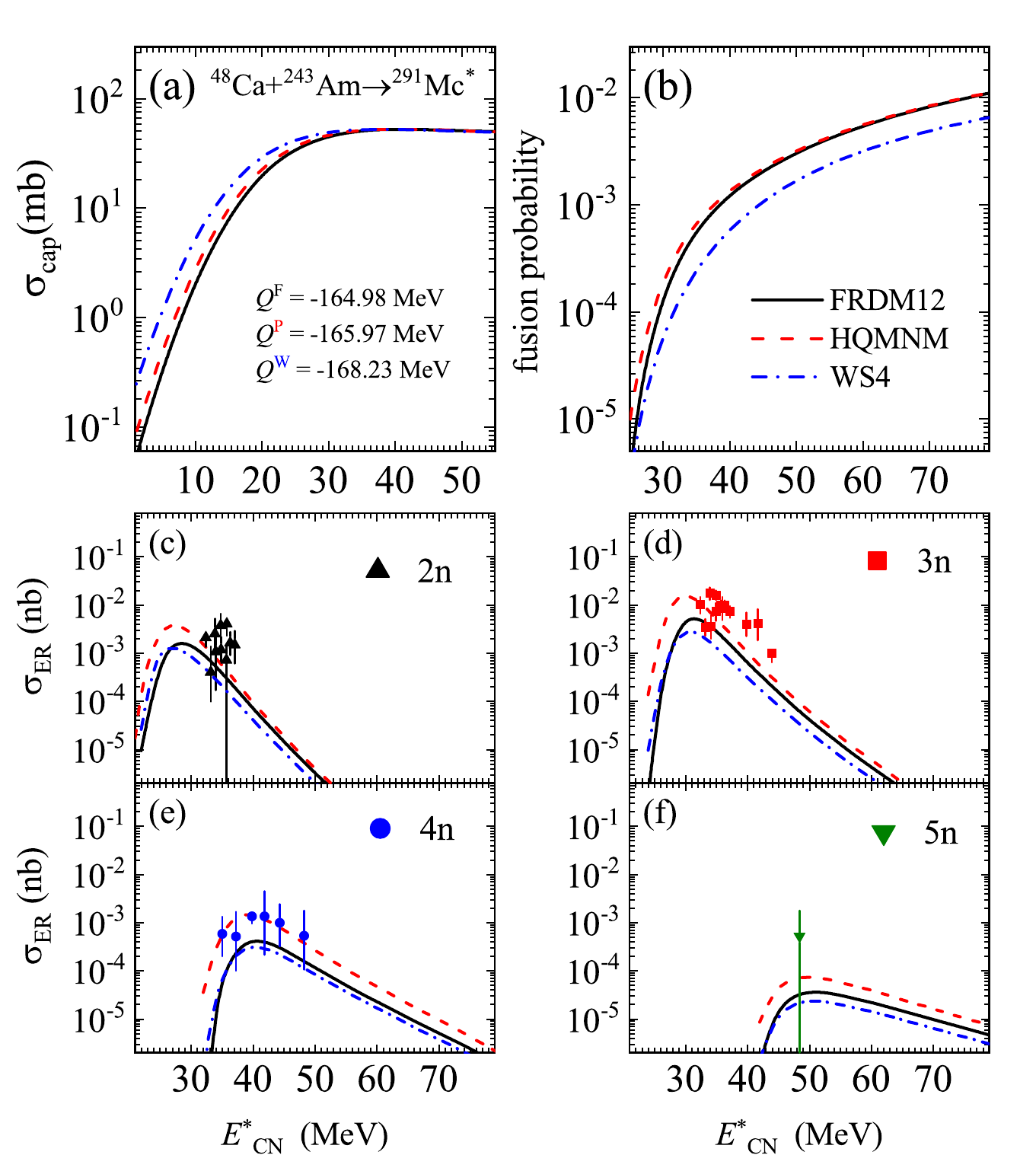}
\caption{\label{fig3}
Panels (a) to (b) depict the capture cross-sections and fusion probabilities for the \(^{48}\text{Ca} + ^{243}\text{Am} \rightarrow ^{291}\text{Mc}^*\). Panels (c)-(f) present the evaporation residue cross-sections  (ERCS) for neutron emission channels ranging from 2 to 5 neutrons, juxtaposed with experimental data. The calculations based on the FRDM12 are represented by solid black lines, while those derived from the HQMNM are denoted by dashed red lines. Experimental ERCS for the neutron channels are symbolized as follows: black up-triangles for the 2-neutron channel, red squares for the 3-neutron channel, blue circles for the 4-neutron channel, and olive down-triangles for the 5-neutron channel.}
\end{figure}

The calculation of the capture probability is independent of nuclear masses. However, different mass tables yield different $Q$ values, which in turn affect the distribution of the capture cross-section with respect to the excitation energy, as shown in Fig. \ref{fig3}(a). The $Q$ values from the three mass tables FRDM12, HFMNM, and WS4 are -164.98 MeV, -165.97 MeV, and -168.23 MeV, respectively. The distribution of the capture cross-section, from left to right, is based on the calculations using the mass tables FRDM12, HFMNM, and WS4. In Fig. \ref{fig3}, the black solid line, red dashed line, and blue dotted line represent the results calculated based on the mass tables FRDM12, HFMNM, and WS4, respectively. The differences in binding energy within various mass tables will affect the calculation of the potential energy surface, resulting in different internal fusion barriers. For the reaction system $^{48}$Ca + $^{243}$Am $\rightarrow$ $^{291}$Mc$^*$, the internal fusion barriers calculated based on the mass tables FRDM12, HFMNM, and WS4 are 9.53 MeV, 9.53 MeV, and 10.41 MeV, respectively. Therefore, we can see from Figure 3(b) that the fusion probability distributions calculated based on the mass tables FRDM12 and HFMNM are very close, while the one based on the WS4 mass table is slightly lower than the other two. It should be noted that the HFMNM mass table only has partial mass information for the superheavy region with Z=110-120, and the masses for other regions are based on experimental data and the FRDM12 mass data. Ultimately, combining the data from the three stages yields the excitation function of the evaporation residue cross-section, whose 2n-5n channel excitation functions are listed in Fig. \ref{fig3}(c)-(f) and compared with experimental data. It can be observed that the calculated results based on the three mass tables are in good agreement with the experimental data, with each differing within an order of magnitude. Among them, the results based on the HFMNM mass table are slightly larger than those based on the other two mass tables, which is due to its larger survival probability.

\begin{figure*}[htb]
\includegraphics[width=0.8\linewidth]{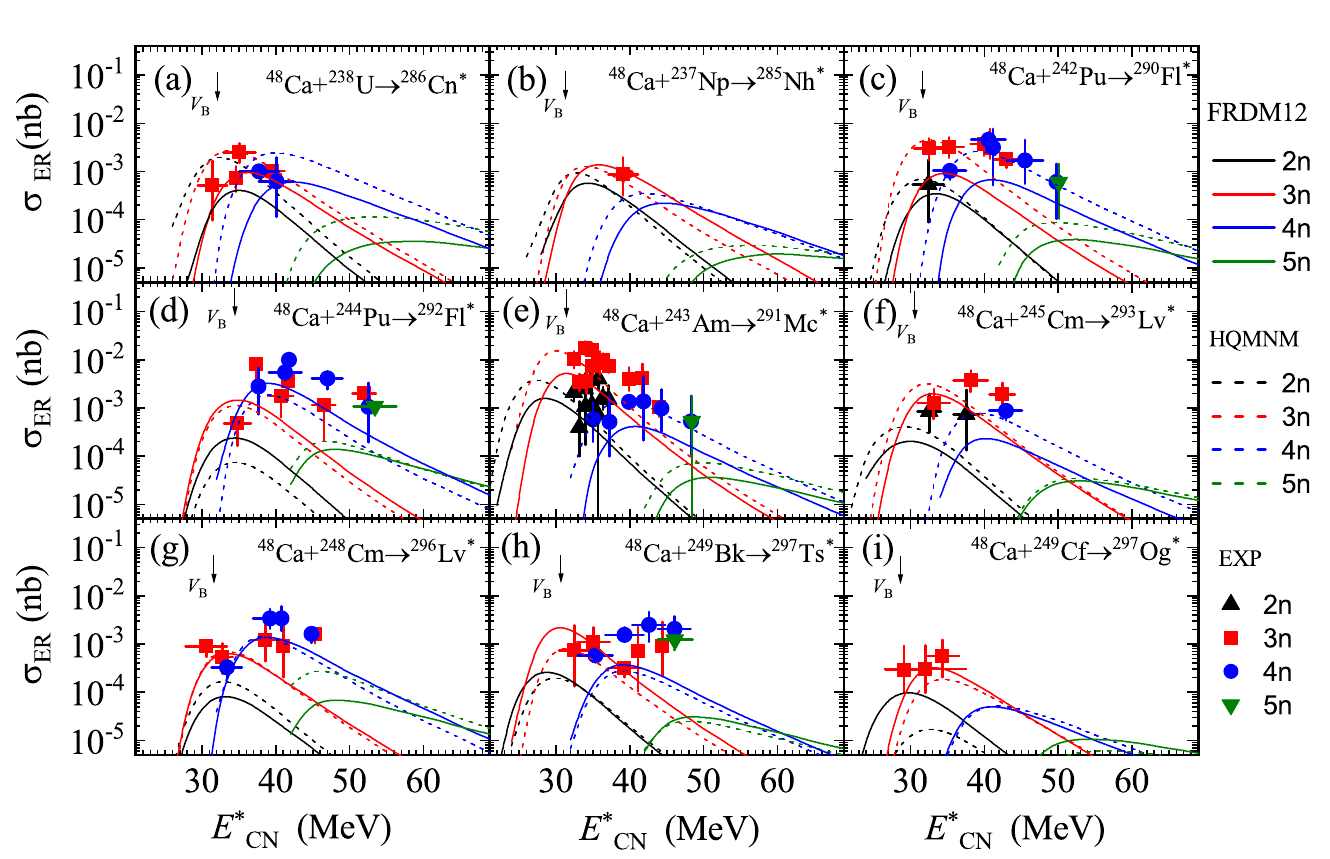}
\caption{\label{fig4} The excitation functions for the ERCS of hot-fusion reactions, resulting in the formation of superheavy elements with atomic numbers \( Z = 112-118 \), are depicted in panels (a) to (i). The theoretical calculations based on the HQMNM are indicated by dashed lines, while those based on the FRDM12 are represented by solid lines, and both are juxtaposed with experimental results for comparison.}
\end{figure*}

Figure \ref{fig3} provides a detailed discussion of the impact that different nuclear mass tables have on the calculations of the three stages—capture, fusion, and survival—in the hot fusion reaction system $^{48}$Ca + $^{243}$Am $\rightarrow$ $^{291}$Mc$^*$. To further validate the performance of the mass tables, we systematically calculated the evaporation residue cross-sections for nine hot fusion reactions ($^{48}$Ca + $^{238}$U/$^{237}$Np/$^{242,244}$Pu/$^{243}$Am/$^{245,248}$Cm/$^{249}$Bk/$^{249}$Cf) based on five mass tables (HQMNM, FRDM12, WS4, KUTY, HFB02) and compared them with experimental data. We list the peak values of the calculated evaporation residue cross-sections and the corresponding peak values of the evaporation residue cross-sections from experimental data in Table \ref{table2}. To clearly demonstrate the influence of mass tables on the calculation of the evaporation residue cross-section excitation function, we only present the calculation results for the mass tables FRDM12 and HQMNM in Fig. \ref{fig4}. It can be seen that there is a good consistency between the calculated results and the experimental results. Overall, the calculation results based on the HQMNM mass table are slightly better than those based on the FRDM12. The relative magnitude between the excitation functions calculated using the two mass tables is not fixed. The HQMNM mass table is constrained by the experimental values of alpha decay energies, and therefore, it can be observed that it conforms well to the experimental results.


\begin{figure*}[htb]
\includegraphics[width=0.85\linewidth]{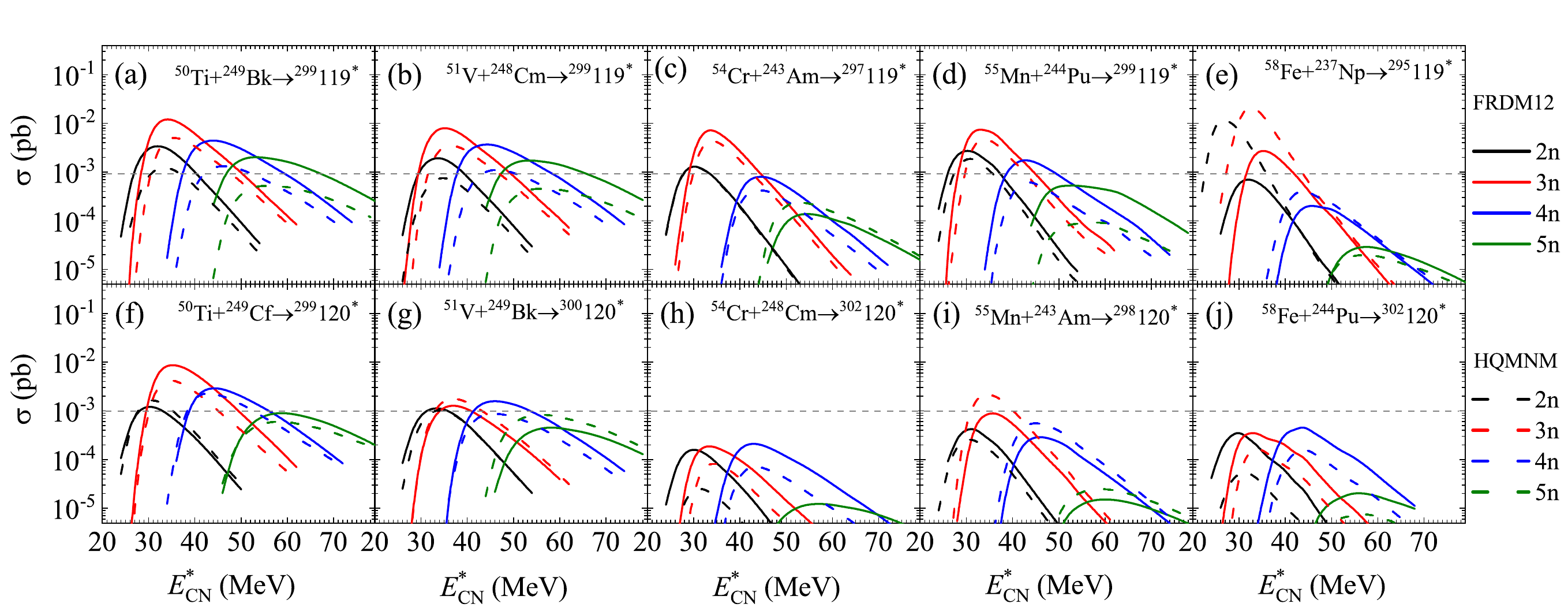}
\caption{\label{fig5} (Color online) Panels show the predicted excitation functions of ERCS for synthesizing superheavy elements Z=119-120 based on the mass tables of HQMNM and FRDM12, which are marked by dash lines and solid lines, respectively.}
\end{figure*}

On the basis of these five masses that can well reproduce existing experimental results, we have continued to calculate ten reaction systems for the synthesis of superheavy elements 119 and 120, namely $^{50}$Cr + $^{249}$Bk, $^{51}$V + $^{248}$Cm, $^{54}$Cr + $^{243}$Am, $^{55}$Mn + $^{244}$Pu, $^{58}$Fe + $^{237}$Np, $^{50}$Ti + $^{249}$Cf, $^{51}$V + $^{249}$Bk, $^{54}$Cr + $^{248}$Cm, $^{55}$Mn + $^{243}$Am, $^{58}$Fe + $^{244}$Pu. The peak values of the evaporation residue cross-sections are listed in Table \ref{table2}. We have listed the results calculated based on the mass tables HFMNM and FRDM12 in Fig. \ref{fig5}. Different colors of lines represent different neutron evaporation channels, and different line styles represent the results of different mass tables. For the synthesis of element 119, the peak distribution range of the five reaction systems based on the HFMNM mass table is from 3.07 to 23.7 fb. For the synthesis of element 120, the peak distribution range of the five reaction systems based on the HFMNM mass table is from 2.8 to 12.8 fb. Except for the $^{58}$Fe + $^{237}$Np system, the results calculated based on the HFMNM are slightly smaller than those calculated based on the FRDM12, which is mainly due to its larger neutron separation energy leading to a smaller survival probability. 

\section{CONCLUSIONS}\label{sec4} 

To accurately predict the synthesis cross-sections of superheavy elements 119 and 120, identify the optimal projectile-target combinations, and determine the optimal collision energies, we have employed the nuclear mass table of HQMNM within the DNS model. Additionally, we compared results with four other four nuclear mass tables. Prior to this, we conducted a detailed discussion on the impact of different mass tables on the capture-fusion-survival stages separately. We found that mass does not affect the capture cross-section, but it does influence the excitation energy corresponding to the capture cross-section. The nuclear mass affects the calculation of the potential energy surface, which in turn causes differences in fusion probabilities. Most importantly, the slight differences in neutron separation energy among various mass tables can significantly affect the magnitude of the survival probability. By considering the impact of nuclear mass on these three processes, one can deduce the effect on the evaporation residue cross-section. Based on the HFMNM mass table, we calculated nine hot fusion reactions and compared them with experimental data, finding that the results are more consistent with the experimental outcomes compared to those of the four other mass tables. On this basis, using the HFMNM mass table, we calculated ten reaction systems for the synthesis of elements 119 and 120, and provided the peak values of the synthesis cross-sections and the corresponding excitation energies, which are listed in Table 2. We discovered that the relative size of the excitation function is primarily determined by the survival probability. In the calculation of the survival probability, we employed the same set of parameters in addition to the nuclear mass table. Therefore, it can be said that the neutron separation energy is crucial for the calculation of the evaporation residue cross-section. For the synthesis of element 119, based on the HFMNM mass table, we found that the cross-section for the $^{58}$Fe + $^{244}$Pu system is the largest, reaching up to 23.7 fb. For the reaction systems of interest, $^{51}$V + $^{248}$Cm and $^{54}$Cr + $^{243}$Am, the maximum cross-sections are 3.6 fb and 4.6 fb, respectively.

\section{ACKNOWLEDGMENTS}\label{sec5}

This work was supported by National Science Foundation of China (NSFC) (Grants No. 12105241,12175072), the Jiangsu Province NSF (Grants No. BK20210788) and the Jiangsu Province University Science Research Project (Grants No. 21KJB140026).


\providecommand{\noopsort}[1]{}\providecommand{\singleletter}[1]{#1}%

\end{document}